# Suppression of stimulated Raman scattering by angularly incoherent light, towards a laser system of incoherence in all dimensions of time, space, and angle

Yi Guo[1], Xiaomei Zhang[1,*], Dirui Xu[1], Xinju Guo[1], Baifei Shen[1,†], Ke Lan[2]

[1]*Department of physics, Shanghai Normal University, Shanghai 200234, China*
[2]*Institute of Applied Physics and Computational Mathematics, Beijing 100094, China*

Abstract:

Laser-plasma instability (LPI) is one of the main obstacles in laser-driven inertial confinement fusion (ICF) for achieving predictable and reproducible fusion at high gain. For the first time we have proved analytically and confirmed with three-dimensional particle-in-cell simulations that angular incoherence has additional and much stronger suppression of the instability growth rate than the well-known temporal incoherence and spatial incoherence usually used in ICF studies. For the model used in our calculations, the maximum field ratio between the stimulated Raman scattering and the driving pulses drops from 0.2 for the Laguerre-Gaussian pulse with a single non-zero topological charge to 0.05 for the super light spring with an angular momentum spread and random relative phases. In particular, angular incoherence does not introduce extra undesirable hot electrons. This opens a novel way to suppress LPI with the light of an angular momentum spread and paves the way towards a low LPI laser system with a super light spring of incoherence in all dimensions of time, space, and angle.

[*] Author to whom correspondence should be addressed. Electronic mail: zhxm@shnu.edu.cn
[†] Author to whom correspondence should be addressed. Electronic mail: bfshen@shnu.edu.cn

Laser-driven plasma instabilities(LPIs) [1-3], such as stimulated Raman scattering (SRS), stimulated Brillouin scattering (SBS), and two-plasmon decay (TPD), is one of the main obstacles in laser-driven inertial confinement fusion (ICF) [4-6] because they may cause significant laser energy loss, generate undesirable hot electrons [5-13], and seriously influence the drive symmetry on the target. To suppress the LPI for ICF, it was proposed to reduce the instability growth rate in the plasma by either the temporal or spatial incoherence of a driving laser. This is because, the incoherence of the laser can be understood as the superimposition of temporal and spatial modes, which changes the amplitude $a$ or intensity $a^2$ of the laser in time or space. We know that the total laser energy can be written as $\propto \int a^2 dt$, while the total instability growth in the linear region can be written as $\propto \int a dt$. Thus, for a given total laser energy, a laser with a changing amplitude exhibits less total instability growth. The effect of incoherence in suppressing the LPI can also be alternatively explained. For two given stimulated waves, such as a plasma wave and a scattering wave, only one mode of the driving laser can couple with them exactly resonantly, while the others contribute less to the instability growth rate. Therefore, for a given average laser intensity, the incoherent laser had a lower instability growth rate. To date, many methods have been proposed to suppress LPI using spatial and temporal incoherence driving lasers, such as laser smoothing techniques [14-18] and broadband laser technology [19-23]. In particular, for the latter, both the theoretical and experimental results indicate that the linear growth of the LPI can be well controlled when the laser bandwidth is much larger than the growth rate [23,24]. Meanwhile, the decoupled broadband lasers [13,25-27] or polychromatic lights composed of multiple colors of beamlets [10,28] show better inhibition effects than continuous broadband lasers.

Even with all these methods, the SRS of the inner laser beams at the National Ignition Facility (NIF) [1,12,29-31] still remains a major issue for high-convergence implosion with a CH capsule inside a gas-filled hohlraum driven by a long laser pulse [32-34].The inner beams of the NIF hohlraums have a long propagation path and encounter plasma blowoff from the high-Z walls, low-Z plasma ablated from the capsule, and any filling gas within the hohlraum, which finally results in the unpredictable, serious SRS. To have a low level of LPI, researchers have to turn to the ablator of high-density carbon (HDC), which has the advantage of high density, allowing the use of thinner capsules and shorter laser pulses, thus leading to a low level of LPI and better symmetry control in low gas-filled hohlraums. With the HDC capsule, the NIF reached the ignition threshold with a target gain of 0.72 in 2021 [11,32]. Nevertheless, the NIF results reveal that the reduction of LPI via both temporal incoherence and spatial incoherence is limited, and the suppression of SRS is still crucial for the NIF towards a predictable and reproducible fusion gain via a high-convergence implosion driven by a long laser pulse.

Note that time and space are related to energy (frequency) and momentum, while the angle is related to the angular momentum. Therefore, it is possible to suppress the LPI via angular incoherence. The spatial coherence area is usually expressed as the product of two space length, that is, $\Delta s = \Delta y \cdot \Delta z$ for a laser propagating in the $x$ direction. We can rewrite this as $\Delta s = r\Delta r\Delta\varphi$. In this way, $\Delta r$ is related to radial incoherence, and $\Delta\varphi$ is related to angular incoherence. Taking SRS as an example, we know that SRS in plasma is a three-wave coupling process referred to the decay of a driving laser into an electron plasma wave and a scattering wave, in which the following frequency and wave number matching corresponding to energy and momentum conservation should be satisfied, $\omega_L = \omega_1 + \omega_2$, $k_L = k_1 + k_2$. It should be noted that in addition to the energy and momentum, the driving laser may also carry angular momentum. Then, angular momentum conservation should also be satisfied for the LPI, that is, $L_L = L_1 + L_2$. In fact, the presence of an orbital angular momentum in the driving laser beam (such as Laguerre-Gaussian beams) has been used to study Raman scattering in laser–plasma interaction [35-38]. In this paper, for the first time we propose to suppress LPI with angularly incoherent light and focus on the effect of angular incoherence on the instability growth rate, which paves the way towards a low-LPI laser system with a super light spring of incoherence in all dimensions of time, space and angle. We prove analytically and with three-dimensional (3D) particle-in-cell (PIC) simulations that angular incoherence has additional and even much stronger suppression of the instability growth rate than temporal incoherence.

The radial and angular momenta of the driving laser can be well described using the following Laguerre–Gaussian modes:

$$a_L(x,t) = a_n \cdot \exp[i\omega_L t - ik_L x + il_L \varphi + \phi], \tag{1}$$

where $a_n = a_0 \cdot (-1)^p [C_{pl}/w(x)] \cdot \left(\sqrt{2}r/w(x)\right)^{|l|} \cdot \exp(-r^2/w(x)^2) \cdot L_p^{|l|}(2r^2/w(x)^2)$ is the transverse profile. Here, $r = \sqrt{y^2 + z^2}$, $\varphi = \tan^{-1}(z/y)$, $w(x) = w_0\sqrt{1 + x^2/x_R^2}$, $R_x = (x^2 + x_R^2)/x^2$, where $a_0$ is the normalized amplitude, $C_{pl}$ is the normalization constant, $L_p^{|l|}(x)$ is the associated Laguerre polynomial, $p$ is the number of radial nodes in the intensity distribution, $k_L = 2\pi/\lambda_L$ is the wave number, $\omega_L$ is the laser frequency, $w_0$ is the beam-waist radius, $l_L$ is the topological charge, and $\phi$ is the original phase at the waist plane. Because in this paper we concentrate on studying the effect of angular incoherence, we set $p \equiv 0$.

Here, we consider a specific case. The driving laser is a superimposition of $N$ modes of different frequency $\omega_L$ and topological charge $l_L$. The laser is written as

$$a_L(x,t) = \sum_{n=1}^{N} a_n \cdot \exp[i\omega_{Ln} t - ik_{Ln} x + il_{Ln}\varphi + \phi_n]. \tag{2}$$

For simplicity, we assume $a_n \equiv a_0$. The frequencies $\omega_{Ln} = \omega_1 + (n-1)\varepsilon_1 \omega_{L0}$ of the modes have the same interval $\varepsilon_1 \omega_{L0}$ and the total bandwidth is $\Delta\omega = (N-1)\varepsilon_1 \omega_{L0}$, where $\omega_1$ is the frequency of the first mode, $\omega_{L0}$ is the central frequency (i.e., corresponding $n = (N+1)/2$) and $\varepsilon_1$ is a constant that indicates the bandwidth gap. Usually, $\Delta\omega/\omega_{L0} \ll 1$. The charges $l_{Ln} = l_1 + (n-1)\varepsilon_2 l_{L0}$ and the total topological charge spread is $\Delta l = (N-1)\varepsilon_2 l_{L0}$, where $l_1$ is the frequency of the first mode, $l_{L0}$ is the topological charge corresponding to the central frequency and $\varepsilon_2$ is a constant that indicates the topological charge gap. Typically, $\varepsilon_2 l_{L0}$ is an integer. If $\phi_n \equiv 0$, the structure of such laser is called as light spring [39,40]. Thus, the distribution of the highest intensities, that is, the strong point, resembles a spring. In this paper, we call a light spring with random phases $\phi_n$ as super light spring. Hereafter, for simplicity, the laser with Laguerre–Gaussian modes is referred to as LG, and the laser with a light spring or super light spring structure is referred to as LS or super LS. Owing to the superimposition, the amplitude for $\phi_n \equiv 0$ can be written as

$$a_L(x,r,\varphi,t) = a_0(r) \frac{\sin\left[\frac{N(\varepsilon_1 \omega_{L0} t - \varepsilon_1 k_{L0} x - \varepsilon_2 l_{L0}\varphi)}{2}\right]}{\sin\left[\frac{\varepsilon_1 \omega_{L0} t - \varepsilon_1 k_{L0} x - \varepsilon_2 l_{L0}\varphi}{2}\right]} \tag{3}$$

This implies that the angular spread is reduced to $2\pi/N$, and the angular position of the highest intensity can be written as $\varphi_{\max} = \varepsilon_1 \omega_{L0} t / (\varepsilon_2 l_{L0})$. By changing $\varphi$ from 0 to $2\pi$, we can obtain the pitch of LS.

The SRS of a non-relativistic laser can be described by [41,42]

$$\left(\frac{\partial^2}{\partial t^2} - c^2 \nabla^2 + \omega_{pe}^2\right) \tilde{\mathbf{a}} = -\tilde{n}_e \mathbf{a}_L, \tag{4}$$

$$\left(\frac{\partial^2}{\partial t^2} + \omega_{pe}^2 - 3v_e^2 \nabla^2\right) \tilde{n}_e = n_0 \nabla^2 (\mathbf{a}_L \cdot \tilde{\mathbf{a}}), \tag{5}$$

where $\omega_{pe}$ is the electron plasma frequency, $\mathbf{a}_L$ is the vector potential of the driving laser pulse, $\tilde{\mathbf{a}}$ and $\tilde{n}_e$ are the vector potential of the back scattering wave and plasma density perturbation, respectively, and $n_0 = \omega_{pe}^2/4\pi e$. We neglect the radial gradient (it is reasonable for a sufficiently large focus spot) and consider that the wave of the form $\sim \exp(i\omega t - ikx + il\varphi)$ is excited. For the scattering wave of lower frequency, we have

$$(\omega^2 - \omega_l^2) = \frac{k^2 c^2 \omega_{pe}^2 a_0^2}{4} \sum_{n=1}^{N} \frac{1}{(\omega - \omega_{Ln})^2 - (k - k_{Ln})^2 c^2 + \frac{(l - l_{Ln})^2}{r^2} c^2 - \omega_{pe}^2} \tag{6}$$

where $\omega_l$ is of the Langmuir wave frequency. Note that the dispersion relation depends on the radial position $r$. For simplicity, we discuss the dispersion relations at radius $R$ of the highest intensity for the LG and LS pulses. Note that only one laser mode is exactly resonant. Denoting the resonant mode as $(\omega_{L0}, k_{L0}, l_{L0})$, we obtain

$$(\omega_l - \omega_{L0})^2 - (k - k_{L0})^2 c^2 + \frac{(l - l_{L0})^2}{R^2} c^2 - \omega_{pe}^2 = 0. \tag{7}$$

Writing $\omega = \omega_l + \delta\omega = \omega_l + i\gamma_s$, with $\delta\omega \ll \omega_l$, we obtain the expression of the instability growth rate $\gamma_s$,

$$\gamma_s \approx \frac{kca_0}{4}\left[\frac{\omega_{pe}^2}{\omega_l(\omega_{L0} - \omega_l)}\right]^{\frac{1}{2}} \sqrt{N}\left(1 - \frac{(N-1)}{8\omega_l}\left(\varepsilon_1 \omega_{L0} + \frac{\varepsilon_2 l_{L0} c}{R}\right)\right)$$

$$\approx \frac{kca_0}{4}\left[\frac{\omega_{pe}^2}{\omega_l \omega_s}\right]^{\frac{1}{2}} \sqrt{N}\left(1 - \frac{(\Delta\omega + \Delta lc/R)}{8\omega_l}\right). \tag{8}$$

where $\omega_s = \omega_{L0} - \omega_l$ is the backward SRS frequency. Here, $\Delta\omega/\omega_{L0} \ll 1$ and $\Delta l/l_{L0} \ll 1$ are assumed, which is reasonable to get the approximate growth rate.

For an actual superimposed laser pulse, the peak laser amplitude is $Na_0$. Therefore, the present formula Eq. (8) for the growth rate is the same as before except for the correction term $(\Delta\omega + \Delta lc/R)/8\omega_l$. The correction term includes two terms that can reduce the instability growth rate. One is the total bandwidth $\Delta\omega$, which has been widely investigated in ICF. The other is the total topological charge spread $\Delta l$. Note that, at $\Delta l > (2\pi R/\lambda_{L0})(\Delta\omega/\omega_{L0})$, the $\Delta l$ term will dominate the correction term and hence the spread of angular momentum plays a more important role than the bandwidth in suppressing $\gamma_s$. Because $\Delta\omega/\omega_{L0} \ll 1$, it is easy to realize $\Delta l > (2\pi R/\lambda_{L0})(\Delta\omega/\omega_{L0})$ for a real superimposed laser pulse.

To confirm the above analysis, we performed 3D PIC simulations using EPOCH code [43]. In all simulations, we set $l_{L0}\varepsilon_2 = 1$. To reduce the simulation time, we take $l_1 = 3$, $N = 7$, resulting in the averaged topological charge $l_{L0} = 6$. Note that a much larger $l_{L0}$ can be used in a real experiment. For simplicity, we consider the same transverse profile for each mode in Eq. (2),

$$a_n \equiv a_0 = a_{00}\left(\frac{\sqrt{2}r}{w_0}\right)^{l_{L0}} \exp\left(-\frac{r^2}{w_0^2}\right).$$

To further reduce the simulation time, we choose a large amplitude $a_{00} = 0.06$, $w_0 = 10$ μm. The central laser wavelength is $\lambda_{L0} = 0.8$ μm (the corresponding central frequency is $\omega_{L0} = 3.75 \times 10^{14}$ HZ). Therefore, the power of the LS is $P = 0.16$ TW. The frequency separation $\varepsilon_1 \omega_{L0} = 0.03 \omega_{L0}$ ($\varepsilon_1 = 0.03$), thus the total frequency spread is $(N-1)\omega_{L0}\varepsilon_1 = 0.18\omega_{L0}$. With this, we obtain the pitch $\Delta x = \frac{2\pi c}{\varepsilon_1 \omega_{L0}} \approx 27$ μm for the LS pulse. The laser has a constant amplitude in time, with a duration of about 93.3 fs. The plasma density in the simulation is $n_e = 1.7 \times 10^{20}$ cm$^{-3}$. The size of the simulation box is 60μm(x)×80μm(y)×80μm(z), corresponding to a moving window of 600×800×800 cells, with one particle per cell. The plasma occupies the 15μm<x<800μm region in the direction of the laser pulse propagation, and -75μm<y<75μm, -75μm<z<75μm. Fig.1(a) shows the 2D *k*-space distribution of the driving laser pulse after propagating 240 μm. As shown, the LS is composed of seven separated frequencies, and the weak signal on the left side of the driving laser is Raman scattering.

For comparison, we consider a broadband LG laser pulse, which is a superimposition of *N* modes with different frequencies but the same topological charge $l_{L0}$, that is, with a spread of $l_{L0}\varepsilon_2 = 0$ for the angular momentum. The total power and other parameters remain the same as those for the above LS pulse. Fig.1(b) shows the 2D *k*-space distribution for the LG case, again after the pulse propagates 240μm. Clearly, there is a much stronger SRS signal in the LG case, which confirms the above theoretical expectation that the $\Delta l$ incoherence dominates the suppression of instabilities. We have also performed the simulations for a laser composed of Gaussian pulses at multiple frequencies, the result of which is similar to the case of the LG pulse of broadband and has less a suppression effect than the LS pulse of $\Delta l$ incoherence.

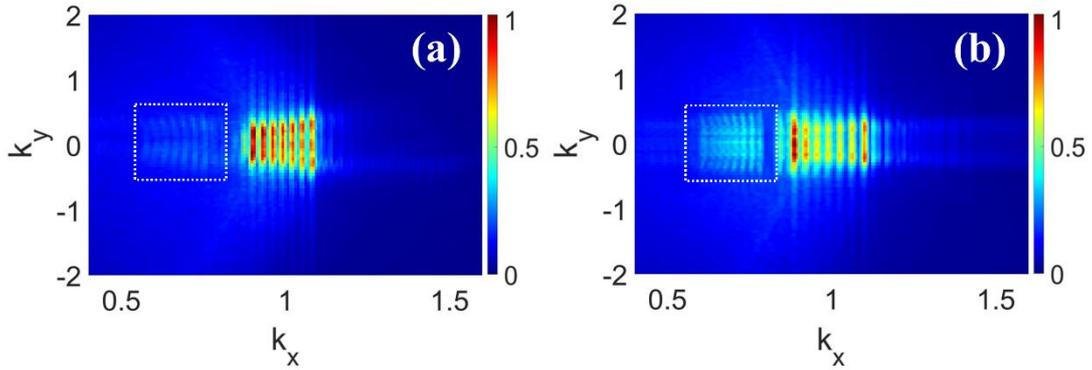

Fig. 1 Spectrum distributions in the *k*-space when the driving laser pulse reaches *x*=240 μm in the case of (a) LS with the topological charge varying from $l = 3$ to $l = 9$ and the total bandwidth is $\Delta\omega = 0.18\omega_{L0}$. (b) LG with the topological charge $l = 6$ and the bandwidth $\Delta\omega = 0.18\omega_{L0}$. The wave number *k* is normalized to $2\pi/\lambda_{L0}$ and the spectrum intensity is normalized to the maximum intensity. The SRS signal is marked with the white dotted box.

To clearly observe the growth process of the SRS due to $\Delta l$, we plot the maximum field ratio between the SRS and the driving pulses along the laser propagation for different $\Delta\omega$ and $\Delta l$ cases in Fig. 2. The maximum field ratio is denoted by $\eta_S = E_s/E_L$, where $E_s$ and $E_L$ are the maximum fields of the SRS and driving pulses, respectively. From Fig. 2, we obtain the following results. First, $\eta_S$ is the highest for the case of $\varepsilon_1 = 0$ and $\varepsilon_2 l_{L0} = 0$, indicating that it has the most serious SRS for the LG pulse with a single frequency and a single topological charge. Second, at $\varepsilon_2 l_{L0} = 0$, i.e., with a single topological charge, $\eta_S$ at 200 μm decreases from 0.36 for the narrowband case of $\varepsilon_1 = 0$ to 0.2 for the broadband case of $\varepsilon_1 = 0.03$. It means that, as well known, the broadband can help to suppress SRS. Third, keeping the same frequency band, we compare the LS case of $\varepsilon_2 l_{L0} = 1$ and the LG case of $\varepsilon_2 l_{L0} = 0$. As shown, for the narrowband case of $\varepsilon_1 = 0$, $\eta_S$ at 200 μm drops from 0.36 of the LG pulse to 0.15 of the LS pulse; and for the broadband case of $\varepsilon_1 = 0.03$, $\eta_S$ drops from 0.2 for the LG pulse to 0.05 for the LS pulse with random phase. These results conceivably prove that the angular spread $\Delta l$ can strongly help suppress SRS growth.

Here we note that if a laser is composed of several beamlets of the same frequency but different topological charges, such as in the case of $\varepsilon_1 = 0$ and $\varepsilon_2 l_{L0} = 1$, then the pitch of the LS is infinite. This implies that the strong point with the highest intensity remains unchanged in the transverse direction. However, as shown in Fig. 2, SRS can still be suppressed to some degree, even though the peak intensity is much larger than that for LG pulse. This strongly supports our idea that the spread of the angular momentum of the laser plays an important role in suppressing the SRS.

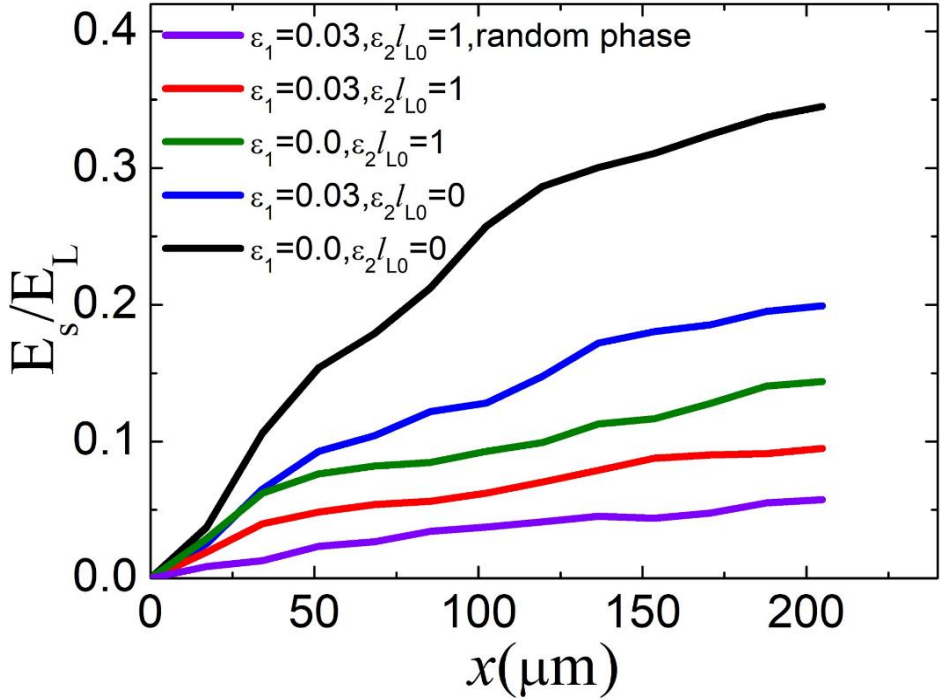

Fig.2. Maximum field ratio $\eta_S = E_s/E_L$ between SRS and driving pulses along the laser propagation for different $\Delta\omega$ and $\Delta l$ cases. The black line indicates the LG case of $\varepsilon_1 = 0$, $\varepsilon_2 l_{L0} = 0$. The blue line indicates the LG case of $\varepsilon_1 = 0.03$, $\varepsilon_2 l_{L0} = 0$. The green line indicates the LS case of $\varepsilon_1 = 0$, $\varepsilon_2 l_{L0} = 1$. The red line indicates the LS case of $\varepsilon_1 = 0.03$, $\varepsilon_2 l_{L0} = 1$. The purple line indicates the super LS case of $\varepsilon_1 = 0.03$, $\varepsilon_2 l_{L0} = 1$ accompanying with the random phases.

One concern is that the superimposition of the LG beamlets produces a strong point in space and time, which may produce extra hot electrons and generates unacceptable shock preheat and entropy in the fuel of the ICF. Thus, it is necessary to investigate whether the LS pulse may arouse more hot electrons and a higher electron temperature than the LG pulse. Presented in Fig.3 is the comparison of the electron distributions in the plasma driven by LG and LS pulses. Here, the plasma wave is still not completely decayed. Obviously, we can see the LS does not produce high temperature nor more hot electrons. Thus, the spread of the angular momentum of the LG pulse does not arouse extra hot electrons.

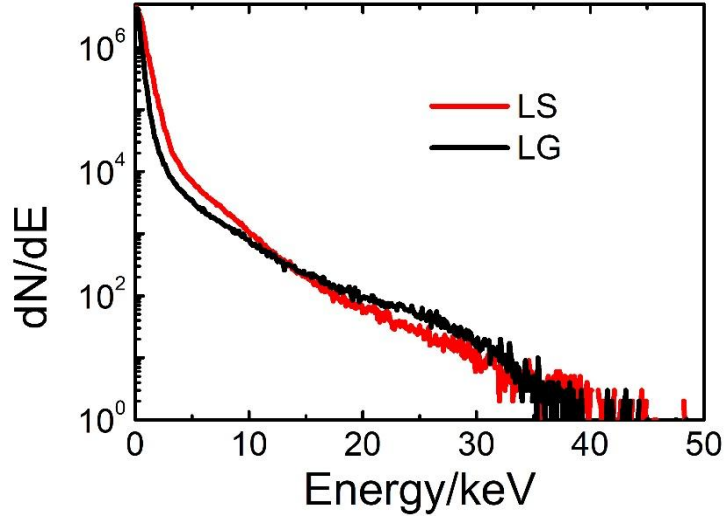

Fig.3. Electron energy spectrum when the driving laser pulse reaches $x=240\mu m$ for the LG pulse with the topological charge $l=6$ and $\varepsilon_1 = 0.03$ (black line) and the LS pulse with the topological charge varying from $l=3$ to $l=9$ and $\varepsilon_1 = 0.03$ (red line).

In a laser facility system, several laser beamlets are usually clustered into a laser bundle, which is used to drive a target. This makes it possible to change the beamlets from usual Gaussian to LG pulses of different topological charge and compose them to a super LS pulse with a spread topological charge. As shown in Fig. 4, LS can be produced by clustering the obliquely incident LG beams of different frequency $\omega_L$ and topological charge $l_L$ owing to the changing thickness of the phase plate azimuthally. The incident angle $\theta$ of each incident LG beam relative to the $x$-axis as shown in Fig.4 differs slightly, which is induced by the small differences between the distances from the beamlets to the center of the source field plane. The pitch of the LS is $\Delta x = 2\pi c/(\omega_{L0}\varepsilon_1)$, with $\omega_{L0}\varepsilon_1 = \Delta\omega/\Delta l$. For large laser facilities, it is still very difficult to control the relative phases of the beamlets. This implies that $\phi_n$ is not the same for different LG pulses. Therefore, we usually obtain a super LS because of the random phase. Although it has several strong point instead of a single perfect strong point, the spread of the angular momentum remains. The lowest growth rate is obtained in this case owing to the additional spatial incoherence, as shown in Fig.2.

In a laser facility system, we can create super LS bundles with incoherence in all dimensions of time, space, and angle by combining LG beamlets. By controlling the phase and frequency of the LG beamlet before combination, we can finally obtain the following three cases of the LS bundle, as shown in Fig. 4: (a) LS light with long pitch, which is narrowband, with one strong point of time-independent intensity; (b) LS light of short pitch, which is broadband, with one strong point of time-dependent intensity; (c) super LS light, with several strong points of time-dependent intensity owing to the random phase. Our results for these three cases confirm that angular incoherence clearly suppress the instability growth rate and the suppression effect of the super LS light is the most significant and is even much stronger than that of the temporal incoherence.

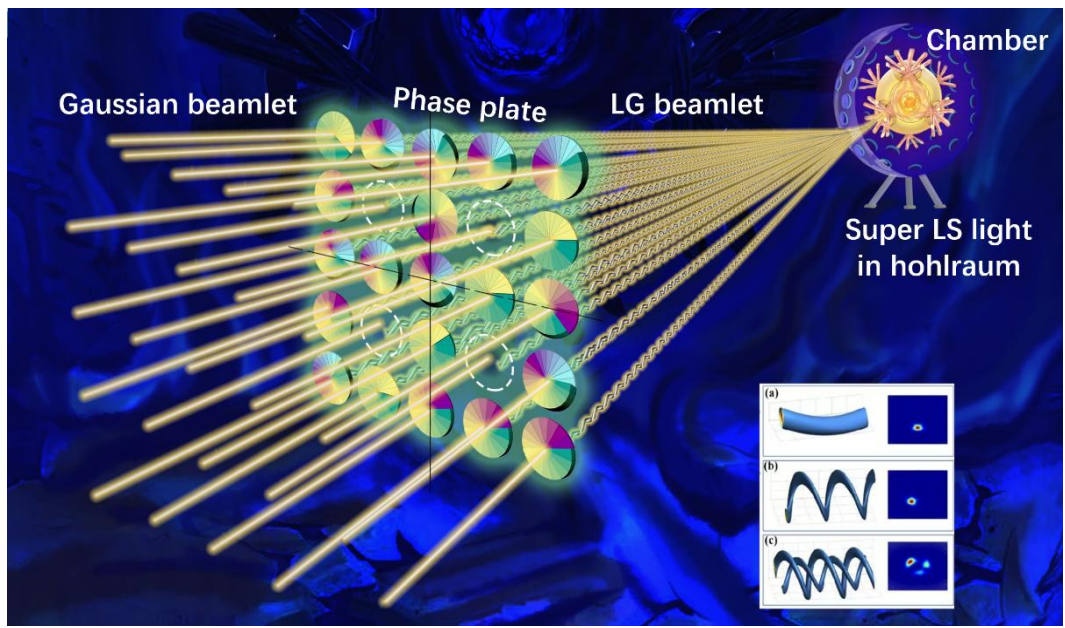

Fig.4. Schematic of generating the super LS light in hohlraum at the center of a target chamber, by combining the LG beamlets at different frequencies, different topological charges, and slightly different incidence angles, which are produced from the Gaussian beamlets by using different phase plates. By controlling the phase and frequency of the LG beamlets via phase plates, we can get the following three cases at the observation plane: (a) LS of long pitch, (b) LS of short pitch, and (c) super LS.

In summary, we have demonstrated the importance of an LS with angular incoherence in suppressing SRS in plasma using both theoretical analysis and 3D PIC simulations, which is of great significance for laser driven ICF. According to our analytical study and 3D PIC simulations, angular incoherence has a much stronger suppression of the instability growth rate than the typically used temporal incoherence. In particular, it is the angular momentum spread that plays an important role instead of the topological charge in suppressing instabilities. In other words, it has little effect on SRS suppression by only increasing the topological charge. In addition, it is interesting to note that the LS pulse does not generate additional hot electrons. We should note in this paper, we consider the suppression of SRS as an example in simulations, but the conclusions can be applied to the suppression of SBS and other parameter instabilities. In a laser system, the laser bundles of a super LS can be generated by combining the LG laser beamlets at different frequencies, different topological charges, and slightly different incidence angles, which are produced from the Gaussian beamlets using different phase plates. Our work opens a novel way to suppress LPI with light of angular incoherence, and a low LPI laser system is expected by using the super LS of incoherence in all dimensions of time, space and angle corresponding to energy (frequency), momentum and angular momentum.

## ACKNOWLEDGMENTS

This work was supported by the National Key R&D Program of China (Grant No. 2018YFA0404803), and the National Natural Science Foundation of China (Grant Nos. 11922515, 11935008, 11335013, and 12035002).